\documentclass[runningheads]{llncs}
\usepackage[T1]{fontenc}
\usepackage{graphicx,verbatim}
\usepackage{multirow}
\usepackage{booktabs}
\usepackage{bbding}
\usepackage{hyperref}
\usepackage[normalem]{ulem}
\useunder{\uline}{\ul}{}
\usepackage[table,xcdraw]{xcolor}
\usepackage{amsfonts}
\usepackage{amsmath}
\usepackage{booktabs}
\usepackage{siunitx}

\usepackage[symbol]{footmisc}

\hypersetup{
   colorlinks,
   linkcolor={green!50!black},
   citecolor={red!50!black},
   urlcolor={blue!80!black}
}
\begin{document}

\title{Single-Subject Multi-View MRI Super-Resolution via Implicit Neural Representations}
\titlerunning{Single-subject Implicit Multi-view Super-resolution for MRI}

\author{Heejong Kim\inst{1}\thanks{Equal contribution.}\thanks{Corresponding author. Email: hek4004@med.cornell.edu} \and
Abhishek Thanki\inst{3}$^{*}$\and
Roel van Herten\inst{1}\and
Daniel J. Margolis \inst{1}\and
Mert R. Sabuncu\inst{1,2}}

\authorrunning{H Kim and A Thanki et al.}

\institute{Department of Radiology, Weill Cornell Medicine, New York, USA\\
\and
School of Electrical and Computer Engineering, Cornell University \\and Cornell Tech, New York, USA \and
Weill Cornell Graduate School of Medical Sciences, New York, USA
}

\maketitle

\begin{abstract}
Clinical MRI frequently acquires anisotropic volumes with high in-plane resolution and low through-plane resolution to reduce acquisition time. Multiple orientations are therefore acquired to provide complementary anatomical information. Conventional integration of these views relies on registration followed by interpolation, which can degrade fine structural details. Recent deep learning–based super-resolution (SR) approaches have demonstrated strong performance in enhancing single-view images. However, their clinical reliability is often limited by the need for large-scale training datasets, resulting in increased dependence on cohort-level priors. Self-supervised strategies offer an alternative by learning directly from the target scans. Prior work either neglects the existence of multi-view information or assumes that in-plane information can supervise through-plane reconstruction under the assumption of pre-alignment between images. However, this assumption is rarely satisfied in clinical settings.
In this work, we introduce Single-Subject Implicit Multi-View Super-Resolution for MRI (SIMS-MRI), a framework that operates solely on anisotropic multi-view scans from a single patient without requiring pre- or post-processing. Our method combines a multi-resolution hash-encoded implicit representation with learned inter-view alignment to generate a spatially consistent isotropic reconstruction. We validate the SIMS-MRI pipeline on both simulated brain and clinical prostate MRI datasets. \textcolor{black}{The code is publicly available for reproducibility\footnote[3]{https://github.com/abhshkt/SIMS-MRI}.}


\end{abstract}
\section{Introduction}
In clinical practice, MRI acquisitions typically provide high in-plane resolution but lower through-plane resolution to reduce scan time, with multiple orientations acquired to offer complementary anatomical views. Integration of these views commonly relies on image registration followed by interpolation~\cite{tsai1984multiframe}, which may blur fine structures, particularly along the low-resolution axis. To address this, super-resolution (SR) methods aim to reconstruct isotropic volumes. Early optimization-based approaches leveraged analytical priors such as low-rank and total variation regularization~\cite{shi2015lrtv}, but struggled under large anisotropy. More recently, deep learning methods trained on paired low- and high-resolution data have demonstrated strong quantitative performance~\cite{pham2019multiscale,oktay2016multi,de2022deep}. However, their clinical applicability is often limited by domain shift, scanner variability, and the absence of standardized MRI intensity scales~\cite{remedios2024pushing}, as well as potential cohort-specific priors.

Starting from SSR~\cite{jog2016self}, self-supervised super-resolution approaches aim to exploit high-resolution information inherently present in the acquired data to enhance the lower-resolution component without relying on external paired data or large-scale training datasets.
ECLARE~\cite{remedios2025ECLARE} models in-plane as the target distribution by leveraging \textcolor{black}{ESPRESO}~\cite{han2023espreso} for slice excitation profile and FOV aware interpolation.
Similarly, SMORE~\cite{zhao2020SMORE} demonstrates that high-resolution in-plane slices can supervise through-plane reconstruction.
However, these methods primarily operate on a single-view anisotropic volume and overlook the fact that MRI data are often acquired in multi-view setups in many clinical domains~\cite{weinreb2016pi,liberman2002breast}. SIMPLE~\cite{benisty2025simple} extends this to join multi-plane adversarial training with per-plane discriminators that use ATME-generated references, but requires a multi-subject training corpus.

More recently, implicit neural representations (INRs) have emerged as a continuous alternative to voxel-based models, providing greater flexibility without being constrained by a discrete voxel grid.
By incorporating Fourier~\cite{tancik2020fourier} and Sinusoidal encoding~\cite{sitzmann2020implicit}, INR based methods have shown promising performance in image reconstruction applications.
In MRI, McGinnis et al.~\cite{mcginnis2023single} demonstrated single-subject multi-contrast super-resolution using INRs with Fourier feature encoding, highlighting the promise of self-supervised continuous models for integrating complementary acquisitions from multi-modal MR images\textcolor{black}{, and IREM~\cite{wu2021irem} applied INR to multi-view MRI reconstruction. However, these frameworks assume pre-registered inputs and do not explicitly account for the inter-scan misalignments that are common in routine clinical acquisitions. NeSVoR~\cite{xu2023nesvor} integrates an implicit volumetric representation with slice-to-volume reconstruction for fetal MRI but accounts only for slice-wise rigid motion.}

In this work, we propose \textbf{S}ingle-subject \textbf{I}mplicit \textbf{M}ulti-view \textbf{S}uper-resolution for MRI (SIMS-MRI), a self-supervised framework that reconstructs isotropic volumes from multi-view anisotropic scans of a single subject without external training data or prior registration. The method follows a three-stage pipeline: (1) hash-encoded implicit image modeling of one view, (2) implicit deformable image registration (IDIR)~\cite{wolterink2022implicit} to align the second view, and (3) multi-view fusion to refine the implicit representation and reconstruct an isotropic volume. We validate the proposed approach on axial and coronal view images simulated brain and clinical prostate MRI datasets, using an orthogonal sagittal view as ground truth for quantitative and qualitative evaluation.

\section{Method}
Our problem formulation considers two anisotropic MRI volumes of the same subject acquired from different views (e.g., axial and coronal view volumes), denoted as $I_{v1}: \Omega \rightarrow \mathbb{R}$ and $I_{v2}: \Omega \rightarrow \mathbb{R}$, where $\Omega \subset \mathbb{R}^3$ represents the spatial domain.
Due to acquisition constraints, both volumes have high-resolution in-plane but thick slices along the through-plane directions.
Our goal is to reconstruct an isotropic high-resolution volume $\hat{I}: \Omega \rightarrow \mathbb{R}$ by leveraging the complementary information available in both views without requiring additional acquisitions. We formulate this as learning a continuous implicit neural representation $f_\theta: \mathbb{R}^3 \rightarrow \mathbb{R}$ that maps a 3D coordinate $\mathbf{x} = (x, y, z)$ to an intensity value, enabling arbitrary-resolution sampling. To handle potential inter-view misalignment, our pipeline learns a deformable registration model $g_\phi: \mathbb{R}^3 \rightarrow \mathbb{R}^3$ that maps coordinates from the one view to the coordinate space of the other.

\subsection{SIMS-MRI}
SIMS-MRI consists of three phases, where we train two INR models sequentially, each with a distinct objective. An image representation modeling, $f_\theta$, is optimized for phases 1 and 3, and an image registration model, $g_\phi$, for phase 2.

\begin{figure}[!htb]
\includegraphics[width=\textwidth]{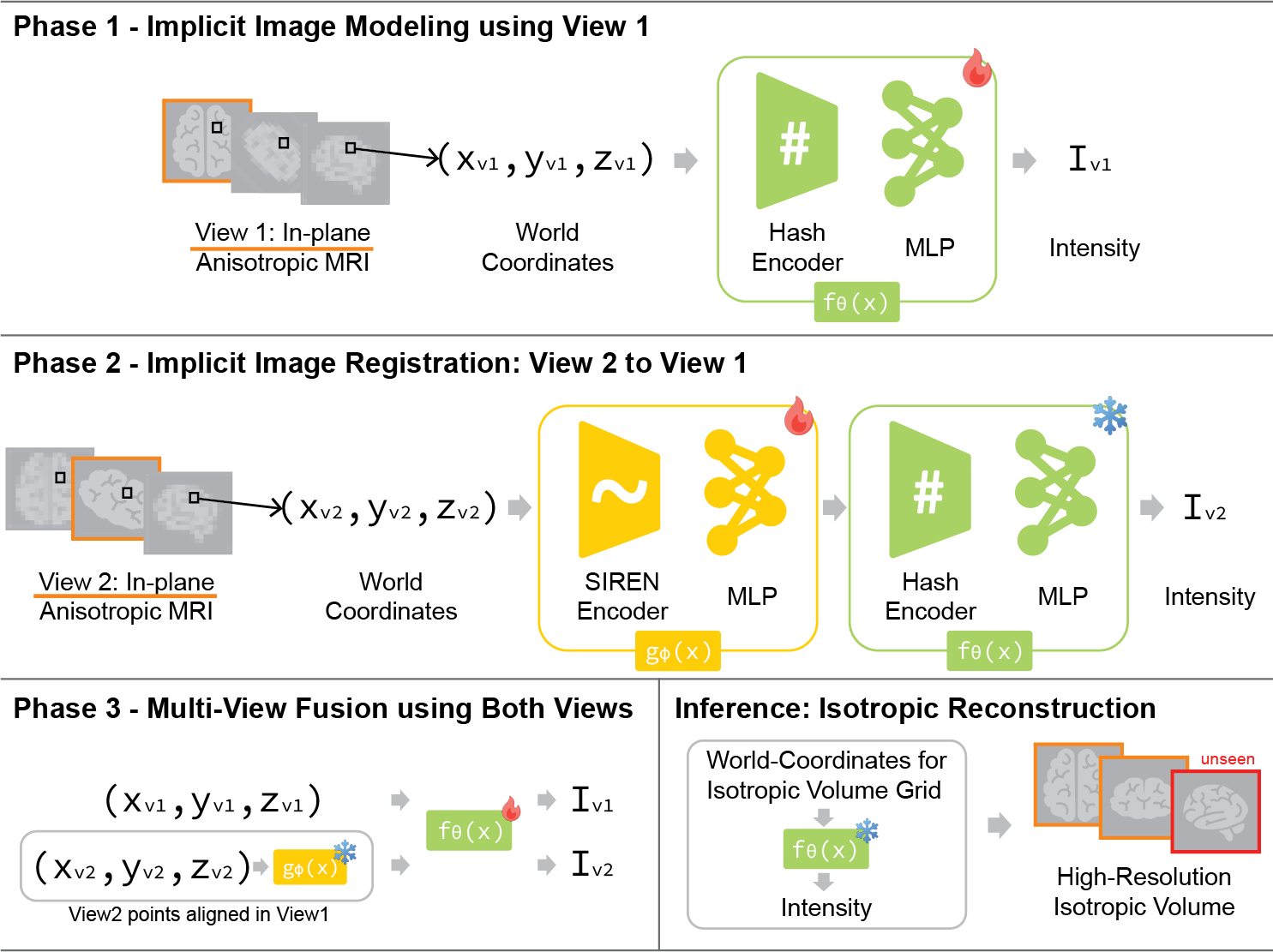}
\caption{
    Overview of the proposed three-phase coordinate-based multi-view MRI fusion framework.
    \textbf{Phase 1:} A continuous implicit image representation model with multi-resolution hash encoding, $f_\theta$, is trained on a reference (axial) view to map 3D world coordinates to voxel intensities.
    \textbf{Phase 2:} A SIREN-based registration network~\cite{wolterink2022implicit} $g_\phi$ learns a dense displacement field that aligns the second (coronal) view to the reference coordinate space while keeping $f_\theta$ fixed.
    \textbf{Phase 3:} The implicit image representation is fine-tuned using both the reference and registered views for information fusion.
    \textbf{Inference}: The learned continuous representation is queried on an isotropic 3D grid to generate a super-resolved volume, enabling reconstruction of unseen slice orientations.
   } \label{fig:overview}
\end{figure}

\subsubsection{Phase 1: Initial Implicit Image Representation.}
The INR-based image representation model, $f_\theta$, is a multi-layer perceptron (MLP) with ReLU activations mapping encoded coordinates to intensity values. The input coordinates are first transformed using a multi-resolution hash encoding before MLP to avoid spectral bias. Specifically, we adopt a coordinate-based INR inspired by Instant-NGP~\cite{muller2022instant}, where multi-resolution hash encoding replaces conventional positional encodings. Unlike fixed projection approaches such as Fourier feature mapping~\cite{tancik2020fourier} and SIREN~\cite{sitzmann2020implicit}, the hash encoding learns adaptive multi-scale embeddings, enabling improved memory efficiency and faster convergence.

\subsubsection{Phase 2: Implicit Image Registration.}
The $f_\theta$ is kept fixed while we train the registration model to align two views. Image registration model, $g_\phi$, predicts a dense displacement field that maps coordinates from the one view to another view's space. Following the IDIR~\cite{wolterink2022implicit}, we use a SIREN network~\cite{sitzmann2020implicit} that maps 3D coordinates to 3D displacements ($\boldsymbol{\delta} = g_\phi(\mathbf{x})$), which are added to the input to obtain the transformed position: $\mathbf{x}' = \mathbf{x} + \boldsymbol{\delta}$. The model was optimized using normalized cross-correlation (NCC)~\cite{lewis1995fast} as the similarity measure. \textcolor{black}{NCC was used over MSE for registration loss because separately acquired MRI views may exhibit intensity-profile differences.}
For the simulated linear movement in two-view brain MRI experiments, we employed a bending energy penalty to encourage near-linear behavior.

\subsubsection{Phase 3: Multi-view Fusion using Both Views.} With the registration model frozen, the pre-trained image model in Phase 1 is trained on both axial and registered coronal views using the same optimizer as phase 1 to learn complementary information of both views.

\subsubsection{Inference: Isotropic Reconstruction.} To obtain the isotropic volume, we query a regular isotropic grid. The grid coordinates are normalized to $[-1, 1]$, consistent with the normalization applied during training. The final isotropic volume is then reconstructed using the original coordinate indices.

\subsubsection{Implementation Details.}
\textcolor{black}{The intrinsic image model consists of a ReLU-activated MLP with a 515-dimensional input layer, four hidden layers of 1024 units each, and a single-unit output layer. A multi-resolution hash encoder has $L=16$ levels, $F=32$ features per level, and a hash table size of $T=2^{19}$ per level. Following IDIR~\cite{wolterink2022implicit}, a SIREN network with three 256-unit hidden layers ($\omega_0 = 32$) is used for the image registration model~\cite{sitzmann2020implicit}. In Phase 2, for the brain dataset, where inter-view differences were assumed to be predominantly rotational, we applied bending energy regularization with a high weighting factor ($\alpha = 1000$) to encourage near-linear transformations~\cite{rueckert1999nonrigid}. With this high weighting factor, the bending energy globally limits the curvature of the deformation field, resulting in a smooth, near-linear transformation. For the prostate dataset, where inter-view differences may include non-linear soft-tissue deformation, no bending energy regularization was applied ($\alpha = 0$). In our current implementation, per-subject optimization requires approximately 30 minutes on a single NVIDIA A100 GPU. The hash grid contains 268.4M trainable embeddings, while the decoder MLP and registration network contain 4.7M and 0.1M trainable parameters, respectively.}

Input images are mapped to scanner coordinates using the affine transformation specified in the image header, and intensity values are jointly normalized to $[-1,1]$.
During training, all phases use batches of $50{,}000$ voxel coordinates. Phases 1 and 3 are optimized using AdamW~\cite{loshchilov2017decoupled} (learning rate $1.2\times10^{-3}$, weight decay $5\times10^{-5}$) with a cosine annealing schedule, while Phase 2 employs Adam~\cite{kingma2014adam} with a learning rate of $10^{-5}$ following IDIR~\cite{wolterink2022implicit}. To mitigate overfitting to high-frequency noise, we adopt a progressive hash-level unlocking strategy following FreeNeRF~\cite{yang2023freenerf}, where hash levels are gradually activated during training to prioritize coarse structures before fine details.

\section{Experiments}
For both experiments, we compare SIMS-MRI against self-supervised super-resolution baselines, including SMORE~\cite{zhao2020SMORE} and ECLARE~\cite{remedios2025ECLARE}, both of which require image registration to integrate information from multiple views. After SMORE and ECLARE reconstruct view-specific isotropic volumes, image registration is applied to align the two views, which are then fused by averaging. In addition, classical interpolation methods, including B-spline and linear interpolation, are included for comparison under the assumption of perfectly aligned inputs. For linear registration, SimpleITK~\cite{lowekamp2013design} was used, and non-linear registration was performed using FireANTs~\cite{jena2024fireants}.

\subsection{Simulated Two-View Brain MRI}
The simulated dataset was generated by $4\times$ downsampling \textcolor{black}{(e.g., 1 mm to 4 mm thickness)} T1-weighted brain MRIs from 50 subjects in the OASIS-3 dataset~\cite{lamontagne2019oasis}. For each subject, two anisotropic volumes were created: an axial volume with high in-plane and low through-plane resolution, and a coronal volume with high in-plane resolution in the coronal plane. To simulate inter-acquisition misalignment, the coronal volume was rotated by $0.1\,\text{rad}$ about the x-axis prior to downsampling. For final isotropic volume of SIMS-MRI, we used isotropic volume grid of ground truth for the inference. An orthogonal view (e.g., sagittal) extracted from the original isotropic volume is used as ground truth for validation. \textcolor{black}{This experiment is designed to isolate anisotropic spatial sampling and inter-acquisition misalignment, while more detailed modeling of slice-selection profiles, slice gaps, and through-plane point-spread functions is left for future work.}

We evaluate the proposed SIMS-MRI framework through comprehensive quantitative and qualitative experiments on a simulated two-view anisotropic 3D brain MRI dataset (e.g., axial and coronal views).
While many recent deep learning approaches emphasize SSIM and PSNR as primary metrics for super-resolution, these measures do not capture perceptual realism~\cite{dohmen2025similarity}.
For quantitative evaluation, we compare pixel-based metrics, including Mean Absolute Error (MAE), Structural Similarity Index Measure (SSIM)~\cite{wang2004image}, and Peak Signal-to-Noise Ratio (PSNR), as well as perceptual metrics such as Learned Perceptual Image Patch Similarity (LPIPS)~\cite{zhang2018unreasonable}, Deep Image Structure and Texture Similarity (DISTS)~\cite{ding2020iqa}, and a Medical Open Network for Artificial Intelligence model-based perceptual loss (MONAI)~\cite{cardoso2022monai} for a more comprehensive assessment. For this perceptual loss, we employed a 3D MedicalNet ResNet-50 model~\cite{chen2019med3d}, to quantify our simulated brain data experiments.

\begin{figure}[!htb]
\includegraphics[width=\textwidth]{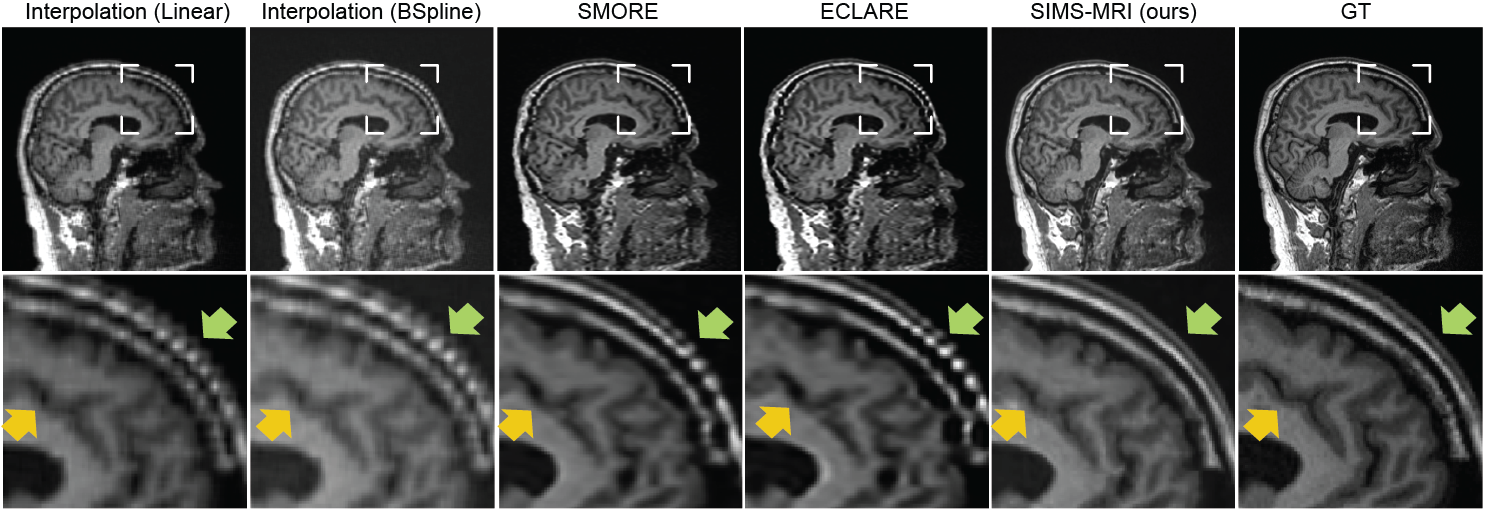}
\caption{Qualitative results for the unseen view in simulated two-view brain MRI. The first row presents a comparison across methods, and the second row shows a zoomed-in view of the highlighted region. Yellow and green arrows indicate areas where SIMS-MRI achieves superior structural detail reconstruction.} \label{fig:result-brain}
\end{figure}

Table~\ref{tab:quant_results} reports quantitative results.
SIMS-MRI achieves the best performance on five out of six metrics and the third-best on the DISTS, demonstrating strong improvements in both intensity-based and perceptual quality measures.
Notably, on MAE, SSIM, and PSNR, SIMS-MRI outperforms competing methods by a substantial margin.
These quantitative gains are reflected in the qualitative results in Fig.~\ref{fig:result-brain}, where SIMS-MRI shows clearer structural delineation and improved intensity consistency compared to baseline methods.
Baseline approaches exhibit noticeable grid artifacts (green arrow), whereas SIMS-MRI produces smoother reconstructions without such artifacts. Furthermore, fine anatomical structures, such as cortical gyri in the sagittal view, are better preserved by SIMS-MRI (yellow arrow), further validating its effectiveness for isotropic MRI reconstruction.

Beyond the comparison with SIMS-MRI, the table also highlights the benefit of multi-view fusion for the baseline methods. Even after image registration, which may introduce slight blurring, both ECLARE-fused and SMORE-fused outperform their single-view counterparts across all metrics, confirming that combining axial and coronal views improves reconstruction quality.

Table~\ref{tab:ablation} presents the ablation results. Replacing hash encoding (HE) with Fourier mapping (FM) substantially degrades performance, underscoring the importance of multi-resolution hash encoding for capturing high-frequency anatomical details. On perceptual metrics (LPIPS, DISTS, and MONAI perceptual score), the SIMS-MRI achieves strong performance, indicating improved structural realism. Incorporating FreeNeRF further enhances distortion-based metrics (MAE, SSIM, and PSNR), improving voxel-wise fidelity while maintaining comparable perceptual quality.

\begin{table}[!htb]
\centering
\small
\caption{Quantitative comparison on simulated two-view brain MRI (unseen view evaluation). '-axial' denotes reconstruction using the axial view only, while '-fused' indicates co-registered and averaged axial/coronal volumes.
($\uparrow$: higher is better and $\downarrow$: lower is better.
Best in \textbf{bold} and second-best in \underline{underline}.)}
\setlength{\tabcolsep}{5pt}
\scalebox{0.9}{
\begin{tabular}{
  l
  S[table-format=1.4]
  S[table-format=1.4]
  S[table-format=1.4]
  S[table-format=2.4]
  S[table-format=1.4]
  S[table-format=1.4]
  S[table-format=1.4]
}
\toprule
{Method} & {MAE $\downarrow$} & {SSIM $\uparrow$} & {PSNR $\uparrow$} & {LPIPS $\downarrow$} & {DISTS $\downarrow$} & {MONAI $\downarrow$} \\
\midrule
Interpolation (B-spline) & 0.0551 & 0.6723 & 18.8914 & 0.2842 & 0.1960 & \num{6.17e-4} \\
Interpolation (Linear)   & 0.0541 & 0.6779 & 19.0907 & 0.2865 & 0.1991 & \num{7.13e-4} \\
ECLARE-axial                & 0.0435 & 0.7616 & 20.5620 & 0.2687 & 0.2040 & \num{3.60e-4} \\
ECLARE-fused          & 0.0412 & 0.7786 & 21.2464 & 0.2464 & \underline{0.1802} & \underline{\num{2.91e-4}} \\
SMORE-axial                 & 0.0431 & 0.7633 & 20.5709 & 0.2700 & 0.2037 & \num{3.57e-4} \\
SMORE-fused           & \underline{0.0409} & \underline{0.7816} & \underline{21.2720} & \underline{0.2445} & \textbf{0.1787} & \num{3.10e-4} \\
SIMS-MRI (ours)           & \textbf{0.0247} & \textbf{0.8877} & \textbf{24.6369} & \textbf{0.2362} & 0.1845 & \sisetup{detect-weight=true}\textbf{\num{2.18e-4}} \\
\bottomrule
\end{tabular}}
\label{tab:quant_results}
\end{table}

\begin{table}[!htb]
\centering
\small
\caption{Ablation study on simulated two-view brain MRI. HE: Hash Encoding; FM: Fourier Mapping. Bold and underlined values indicate the best and second-best results.}
\setlength{\tabcolsep}{5pt}
\scalebox{0.9}{
\begin{tabular}{
  l
  S[table-format=1.4]
  S[table-format=1.4]
  S[table-format=1.4]
  S[table-format=2.4]
  S[table-format=1.4]
  S[table-format=1.4]
  S[table-format=1.4]
}
\toprule
{Method} & {MAE $\downarrow$} & {SSIM $\uparrow$} & {PSNR $\uparrow$} & {LPIPS $\downarrow$} & {DISTS $\downarrow$} & {MONAI $\downarrow$} \\
\midrule
SIMS-MRI               & \textbf{0.0247} & \textbf{0.8877} & \textbf{24.6369} & \underline{0.2362} & \underline{0.1845} & \underline{\num{2.18e-4}} \\
w/o FreeNeRF \& w/ HE             & \underline{0.0281} & \underline{0.8461} & \underline{23.6493} & \textbf{0.2254} & \textbf{0.1744} & \sisetup{detect-weight=true}\textbf{\num{1.64e-4}} \\
w/o FreeNeRF \& w/ FM              & 0.0327 &  0.8077 & 23.3116 & 0.3577 & 0.2345 & \num{5.31e-4} \\
\bottomrule
\end{tabular}}
\label{tab:ablation}
\end{table}

\subsection{Real-world Clinical Prostate MRI}
For further evaluation, we assessed our method using routinely acquired clinical prostate MRI images from 50 patients at a single hospital, which typically are acquired in $0.5\times0.5mm^2$ in-plane with $3mm$ slice thickness. The acquisition series included three in-plane acquisitions, resulting in axial, coronal, and sagittal volumes.
\textcolor{black}{We used the anisotropic axial and coronal volumes as inputs and the subsampled isotropic grid for the final isotropic inference grid.
The sagittal volume was used as a held-out pseudo-reference for evaluation. Because it was not explicitly co-registered with the reconstructed volumes, it provides only a weak reference for evaluating consistency in an unseen orientation rather than an exact voxel-wise ground truth. To avoid potential blurring artifacts that volume registration can introduce, we performed slice-wise comparison based on spatial proximity rather than voxel-level alignment. This procedure was applied consistently to all baseline methods as well as SIMS-MRI.}

The MONAI perceptual metric, which is robust to minor spatial misalignment as other perceptual metrics and better reflects structural consistency in medical images, was computed on the sagittal images by matching each reconstructed slice to the closest corresponding slice in physical coordinate space. A 2D ResNet-50 model pretrained on RadImageNet~\cite{mei2022radimagenet} was used for prostate MRI experiments.

\begin{figure}[!htb]
\includegraphics[width=\textwidth]{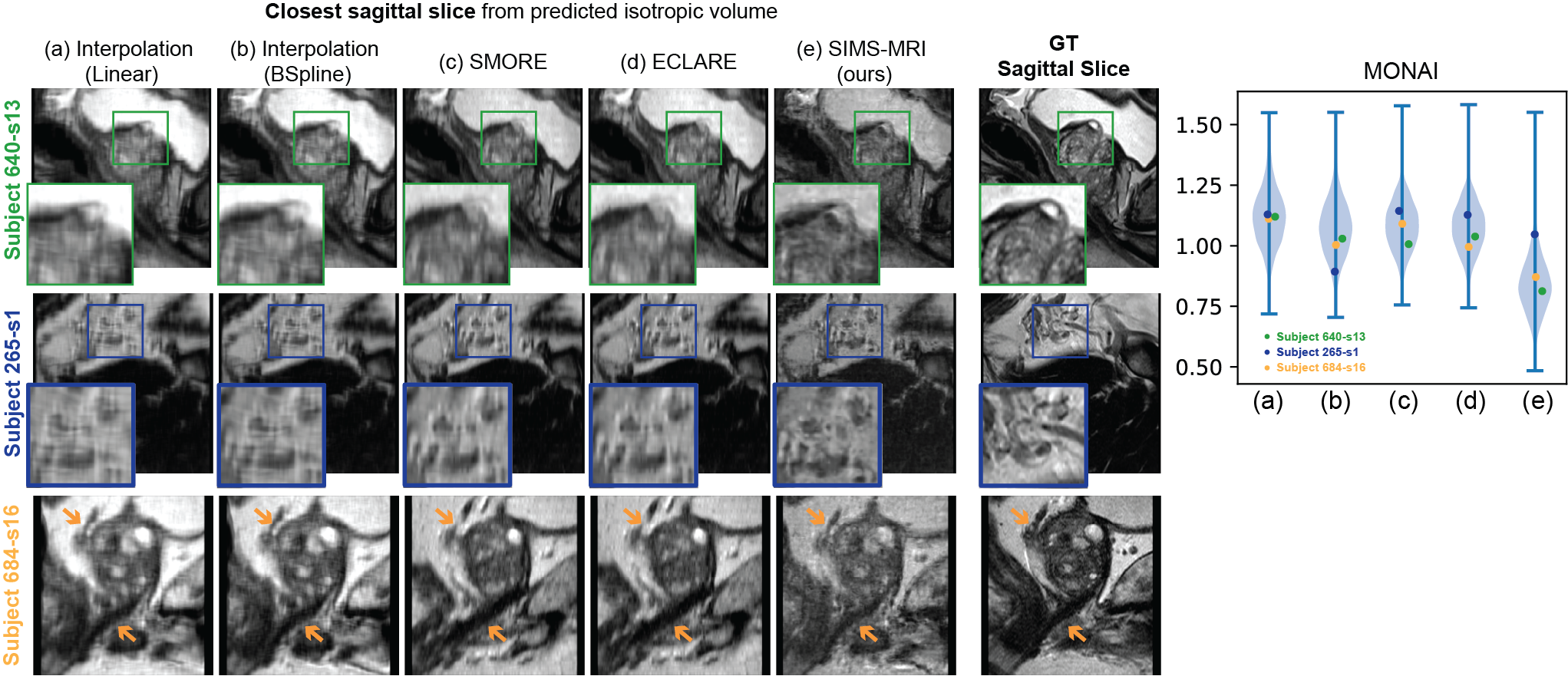}
\caption{Results for prostate MRI. The violin plots show the MONAI metric (MONAI: Medical Open Network for Artificial Intelligence model-based perceptual metric). Each row corresponds to a representative subject, selected from slices near the median of the distribution shown in the violin plots (color-coded and marked in the plots).} \label{fig:result-prostate}
\end{figure}

Figure~\ref{fig:result-prostate} summarizes the qualitative and quantitative comparison between baseline methods and the proposed SIMS-MRI. Compared to baseline methods (a–d), SIMS-MRI (e) exhibits a clear downward shift in the distribution, indicating lower perceptual discrepancy in the unseen sagittal view. The highlighted representative subjects lie close to the median of each distribution, demonstrating that the qualitative examples shown on the left are representative rather than selectively chosen. In the first row, SIMS-MRI achieves improved structural similarity, with clearer anatomical boundaries and better preservation of spatial organization. The second row highlights improved textural consistency, where SIMS-MRI better maintains fine-grained intensity patterns. The third row demonstrates simultaneous improvements in both structural and textural fidelity.

\section{Discussion and Conclusion}
We presented SIMS-MRI, a single-subject, self-supervised implicit framework for multi-view MRI super-resolution that jointly models continuous image representation and inter-scan misalignment without external training data or prior registration. By combining multi-resolution hash encoding with implicit deformable alignment and multi-view refinement, the method integrates complementary anisotropic acquisitions into a spatially consistent isotropic reconstruction, reducing sensitivity to domain shift, scanner variability, and cohort-level priors. While the per-subject optimization is more computationally demanding than feed-forward approaches and clinical evaluation was limited to slice-level correspondence, SIMS-MRI consistently improved quantitative performance in simulated data and perceptual quality in real clinical prostate MRI. \textcolor{black}{Future work will extend this framework to additional views and contrasts, evaluate robustness across different anisotropy ratios, and improve the efficiency of the pipeline for practical deployment. Overall, we believe SIMS-MRI represents a meaningful step toward data-efficient isotropic MRI reconstruction from complementary anisotropic MRI acquisitions.}

\begin{credits}
\subsubsection{\ackname} \textcolor{black}{Funding for this project was in part provided by the NIH grants R01AG053949, R01AG064027, R01AG070988, R01HL174863, U54DK144866, K25CA283145, and the NSF CAREER 1748377 grant. Parts of Figure~\ref{fig:overview} were designed using icons from Flaticon.com.}

\subsubsection{\discintname} \textcolor{black}{The authors have no competing interests to declare that are relevant to the content of this article.}
\end{credits}

\bibliographystyle{splncs04}
\bibliography{bib}
\end{document}